\documentclass[aps,twocolumn,superscriptaddress,showpacs,preprintnumbers,amsmath,amssymb]{revtex4-2}
\usepackage[utf8]{inputenc}
\usepackage[dvipsnames]{xcolor}
\usepackage[normalem]{ulem}
\usepackage{comment}
\usepackage{overpic}
\usepackage{multirow}
\usepackage{booktabs} 
\usepackage{amsmath}
\usepackage[colorlinks,linkcolor=red,anchorcolor=blue,citecolor=blue,urlcolor=blue,breaklinks=true]{hyperref}
\usepackage{cleveref}
\newsavebox{\tablebox}
\usepackage{graphicx}
\usepackage{dcolumn}
\usepackage{bm}
\usepackage{gensymb}
\usepackage{tabularx}
\usepackage{booktabs}
\usepackage{multirow}
\usepackage{lipsum}
\usepackage{subfigure}
\usepackage{float}
\usepackage{lineno} 
 \newcommand{\BESIIIorcid}[1]{\href{https://orcid.org/#1}{\hspace*{0.1em}\raisebox{-0.45ex}{\includegraphics[width=1em]{orcidIcon.pdf}}}} 

\usepackage{orcidlink}
\definecolor{lime}{HTML}{A6CE39}
\DeclareRobustCommand{\orcidicon}{
	\begin{tikzpicture}
	\draw[lime, fill=lime] (0,0) 
	circle [radius=0.2] 
	node[white] {{\fontfamily{qag}\selectfont \tiny ID}};
	\draw[white, fill=white] (-0.0625,0.095) 
	circle [radius=0.007];
	\end{tikzpicture}
	\hspace{-2mm}
}

\foreach \x in {A, ..., Z}{\expandafter\xdef\csname orcid\x\endcsname{\noexpand\href{https://orcid.org/\csname orcidauthor\x\endcsname}
			{\noexpand\orcidicon}}
}

\begin{document}

\title{\boldmath  
Dalitz decay of $K^*(892) \rightarrow  K \ell^+\ell^-$: A New Probe for Hadronic Structure and Dark Photon Searches
}

\author{Benhou Xiang~\orcidlink{0009-0001-6156-1931}}\email{xiangbh@ihep.ac.cn}
\affiliation{  Institute of High Energy Physics,
                  Chinese Academy of Sciences, Beijing  100049, China}
\affiliation{ University of Chinese Academy of Sciences,
                  Beijing  100049, China}
\author{Wanling Chang~\orcidlink{0000-0001-9985-9402}}\email{changwl@ihep.ac.cn}
\affiliation{ Henan University of Animal Husbandry and Economy,
                  Zhengzhou 450046, China}
                  
\author{Shuangshi Fang~\orcidlink{0000-0001-5731-4113}}\email{fangss@ihep.ac.cn}
\affiliation{  Institute of High Energy Physics,
                  Chinese Academy of Sciences, Beijing  100049, China}
\affiliation{ University of Chinese Academy of Sciences,
                  Beijing  100049, China}
\affiliation{      Center for High Energy Physics, Henan Academy of Sciences, Zhengzhou 450046, China
                  }
\author{Jingqing Zhang~\orcidlink{0000-0003-3314-2534}}\email{zhangjq@ihep.ac.cn}
\affiliation{  Nanjing Normal Univerisity, Nanjing 210097, China}

\begin{abstract}

We present the first comprehensive study of the rare Dalitz decay $K^*(892) \rightarrow K \ell^+ \ell^- (\ell = e, \mu)$, providing a prediction for the  branching fraction and the dilepton mass spectrum.  This decay involves the emission of a virtual photon which  converts into a lepton pair, offering a probe of the transition form factor $F_{K^*K}(q^2)$ and underlying meson structure. Using a single pole approximation for the form factor, we present the  calculation of the branching fraction for this rare decay channel for the first time. Furthermore, we also investigate the potential to search for a light $A^\prime$ boson (dark photon) appearing as a narrow resonance in the dilepton spectrum, and discuss the experimental sensitivity and new physics opportunities at the dedicated BESIII experiment. Our results establish $K^*(892) \rightarrow K \ell^+ \ell^- (l = e, \mu)$ as a new laboratory for hadronic structure and dark-sector searches.

\end{abstract}

\maketitle

\section{Introduction}
Electromagnetic (EM) Dalitz decays of vector mesons to pseudoscalar mesons and lepton pairs, such as $V \to P l^+ l^-$, offer a unique window into the fundamental processes governing hadronic structure and photon-meson interactions. Historically, these processes have provided a unique window into the non-perturbative regime of Quantum Chromodynamics (QCD), enabling the extraction of dynamic electromagnetic transition form factors that encode the internal structure of hadrons~\cite{Landsberg:1985kjr,Landsberg:1985gaz}.
Besides the decays like $\omega\to\pi^0\ell^+\ell^-$, $\phi\to\eta\ell^+\ell^-$, more recently charmonium transitions such as $J/\psi \to P\ell^+\ell^-$ have been extensively investigated both theoretically and experimentally, revealing critical insights into dynamic transition form factors and the nature of the intermediate virtual photons responsible for lepton pair production~\cite{Fu:2011yy,BESIII:2025xjh,He:2020jvj,BESIII:2018qzg,BESIII:2018aao,BESIII:2014dax}. However, much less attention has been paid to the analogous decays involving strange vector mesons, particularly the  the lightest strange vector $K^*(892)$, despite their unique role in flavor physics.

The $K^*(892)\rightarrow K l^+ l^-$, mediated by an intermediate virtual photon,  provides direct access to the electromagnetic transition form factor $F_{K^*K}(q^2)$, which encodes the structure of the $K^* \to K\gamma^*$ vertex as a function of the timelike momentum transfer $q^2$. Precise measurement of the $q^2$-dependence of this form factor offers a crucial test of theoretical models, including Vector Meson Dominance (VMD), chiral perturbation theory extended with vector resonances, and dispersive approaches. 
It reveals potential contributions from higher-mass resonances and constrains hadronic inputs crucial for other precision observables. Furthermore, comparison between the electron and muon channels probes lepton universality and the effects of lepton mass in a radiative transition.

Additionally, the Dalitz decays provide a clean experimental environment to search for weakly coupled light vector boson, the so-called dark photon($A^\prime$)~\cite{Fabbrichesi_2021}, that kinetically mixes with the Standard Model photon. In many minimal dark-sector models the $A^\prime$  acquires direct couplings to charged leptons proportional to a small mixing parameter $\varepsilon$ and can be produced as a real or virtual state in meson decays. If a light dark photon exists with mass in the MeV--GeV range, it would manifest as a narrow resonance in the dilepton invariant-mass spectrum superimposed on the continuum from the conventional Dalitz decay. The channel $K^*(892)\rightarrow K \gamma^* \rightarrow K \ell^+\ell^-$,
is particularly attractive because the parent meson is a vector state that couples directly to the electromagnetic current, enabling sensitive production of an $A^\prime$  through mixing, and because strange-meson decays probe coupling structures and mass ranges that are complementary to searches in $\pi^0$, $\eta$ and heavy-flavor decays.

Experimentally, modern  $e^+e^-$ colliders (e.g., BEPCII/BESIII~\cite{Yu:2016cof,BESIII:2009fln}) and hadronic facilities (e.g., LHCb~\cite{LHCb:2008vvz,LHCb:2014set}) can produce large samples of  $K^* (892)$ resonances, enabling statistically powerful studies of rare Dalitz decays and searches for narrow resonant structures in dilepton spectra. The relatively simple final state $K^*(892)\rightarrow K \ell^+\ell^- $  facilitates reconstruction and background suppression, while the expected  Dalitz rate provides a well-understood normalization for setting limits on or potentially discovering a dark photon. A dedicated analysis could therefore sharpen constraints on the kinetic-mixing parameter
across a dark-photon mass range that is otherwise challenging, and, in the event of a signal, yield immediate implications for the dark-sector model.  Most recently, a proposal for Super Charm-Tau Facility (STCF)~\cite{Achasov:2023gey}
was presented, which will accumulate unprecedented data samples containing large samples of $K^*(892)$ resonances. 

Motivated by recent experimental progress, a fresh theoretical analysis of $K^*(892) \to K\ell^+ \ell^-$ is timely. The decay process is analogous to well-studied channels like $J/\psi \to P \ell^+ \ell^-$, for which the analytic decay rate and angular distributions have been derived using the vector dominance model and the pole approximation for the transition form factor. Applying these tools to the $K^*$ system allows us to predict the branching fraction, understand the underlying hadronic physics, and provide benchmarks for experimental dark photon searches. Furthermore, the structure expected in the dilepton invariant mass spectrum could illuminate subtle effects from QCD as well as possible footprints of physics beyond the Standard Model. In this work, we present the first calculation of the $K^*(892) \to K\ell^+ \ell^-$ branching fraction, detail the methodology and theoretical assumptions, and analyze the prospects of observing a dark photon in the lepton pair spectrum, laying a firm foundation for future experimental investigations.

\section{Branching fraction of $K^*(892) \to K \ell^+\ell^-$}

Following the vector meson to pseudoscalar meson  Dalitz decay formalism summarized in Ref.~\cite{Fu:2011yy}, normalizing the dileptonic mode to the corresponding radiative decay and integrating the point-like QED spectrum times the transition form factor, one can obtain the $q^2$-dependent differential decay width, 

\begin{widetext}

\begin{eqnarray}
 \frac{d\Gamma(K^*\rightarrow K \ell^+\ell^-)}{d q^2} =\frac{1}{3}
 \frac{\alpha^2}{24\pi m_{K^*}^3} \frac{|f_{K^*K}(q^2)|^2}{q^2}\left(1-\frac{4m^2_\ell}{q^2}\right)^{1/2}
\left(1+\frac{2m^2_\ell}{q^2}\right)
 \left[(m^2_{K^*}-m^2_K+q^2)^2-4m^2_{K^*}q^2\right]^{3/2},
 \label{eq:dgamma}
\end{eqnarray}
\end{widetext}
where, $\alpha$ is the fine-structure constant, $m_l$ is the lepton mass,  $q^2$ is the invariant mass squared of the lepton pair ($4m_\ell^2 \leq q^2 \leq (M_V - M_P)^2$). $m_{K^*}$ and $m_K$ are the masses of the initial $K^*(892)$
state and pseudoscalar meson.
Throughout this work, we distinguish two related quantities: the unnormalized transition form factor (TFF), denoted by $f_{K^*K}(q^2)$, and the normalized form factor, defined as $F_{K^*K}(q^2)\equiv f_{K^*K}(q^2)/f_{K^*K}(0)$ with the convention $F_{K^*K}(0)=1$. The differential width in Eq.~(\ref{eq:dgamma}) is expressed in terms of the unnormalized TFF $f_{K^*K}(q^2)$, whereas the VMD parametrization in Eq.~(\ref{eq:vmd_form}) and the normalized ratio in Eq.~(\ref{eq:dgamman}) are formulated in terms of the normalized form factor $F_{K^*K}(q^2)$. The physical electromagnetic transition form factor, as would be extracted from experimental measurements, is the normalized quantity $F_{K^*K}(q^2)$, since the absolute normalization $f_{K^*K}(0)$ is absorbed into the radiative decay width $\Gamma(K^*\to K\gamma)$.

We employ the Vector Meson Dominance (VMD) model to describe this form factor. In VMD, the virtual photon ($\gamma^*$) couples to the hadronic vertex through intermediate vector meson states. For the $K^* \to K \gamma^*$ transition, the dominant contributions are expected from the light vector mesons $\rho^0$, $\omega$, and $\phi$, which mix with the photon. The form factor can be expressed as a sum of pole terms:
\begin{equation}
F_{K^*K}(q^2) = \sum_{V=\rho,\omega,\phi} \frac{a_V m_V^2}{m_V^2 - q^2 - i m_V \Gamma_V},
\label{eq:vmd_form}
\end{equation}
where $m_V$ and $\Gamma_V$ are the mass and width of the intermediate vector meson $V$, and $a_V$ are coefficients encoding the relative contributions of each vector meson to the $K^*K$ transition. These coefficients satisfy the normalization condition $\sum_V a_V = F_{K^*K}(0) = 1$, which follows from the definition of the normalized form factor. The general multi-resonance expression in Eq.~(\ref{eq:vmd_form}) is given for completeness; a full determination of the individual $a_V$ would require information from SU(3) flavor symmetry and the known radiative width $\Gamma(K^*\to K\gamma)$, which is deferred to future work. For our numerical evaluation in this paper, we adopt the \textit{single-pole approximation}, retaining only the dominant $\rho$ contribution. This corresponds to setting $a_\rho=1$ and $a_\omega=a_\phi=0$, giving the simplified form
\begin{equation}
F_{K^*K}(q^2) \approx \frac{m_\rho^2}{m_\rho^2 - q^2}.
\label{eq:pole}
\end{equation}
The resulting expression is consistent with the classic pole approximation employed in Ref.~\cite{Fu:2011yy,BESIII:2014dax} for light meson systems.

For the corresponding radiative decay of $K^* \rightarrow K
\gamma$, the decay width can be obtained as:
\begin{eqnarray}
\Gamma(K^*(892) \rightarrow K \gamma ) = \frac{1}{3}
\frac{\alpha(m^2_{K^*} - m^2_K)^3} {8 m^3_{K^*}}|f_{K^* K}(0)|^2.
 \label{eq:gammap}
\end{eqnarray}

From Eqs.~(\ref{eq:dgamma}) and~(\ref{eq:gammap}) the $q^2$-dependent differential decay width in the $K^* \rightarrow K\ell^+\ell^- $ decay normalized to the width of the corresponding radiative $K^*\rightarrow  K\gamma $ is derived to be 

\begin{widetext}
\begin{eqnarray}
 \frac{d\Gamma(K^* \rightarrow K l^+l^-)}{d q^2 \Gamma (K^* \rightarrow K \gamma)}
 &=&
 \frac{\alpha}{3\pi} \left|\frac{f_{K^*K}(q^2)}{f_{K^* K}(0)}\right|^2 \frac{1}{q^2} \left(1-\frac{4m^2_l}{q^2}\right)^{1/2}
\left(1+\frac{2m^2_l}{q^2}\right)
 \left[\left(1+\frac{q^2}{m^2_{K^*}-m^2_K}\right)^2 - \frac{4m^2_{K^*}q^2}{(m^2_{K^*}-m^2_K)^2}\right]^{3/2}\nonumber \\
 &=& |F_{K^*K}(q^2)|^2 \times [\mbox{QED}(q^2)],
 \label{eq:dgamman}
\end{eqnarray}
\end{widetext}

where the normalized form factor for the $K^* \rightarrow K\gamma^*$ transition is defined as $F_{K^*K}(q^2)\equiv f_{K^* K}(q^2)/f_{K^* K}(0) $, and the normalization is defined such that $F_{K^*K}(0) =1$. The form factor defines the electromagnetic properties of the region in which $K^*$ is converted into pseudoscalar. By comparing the measured spectrum of the lepton pairs in the Dalitz decay with QED calculations for point-like particles, it is possible to determine
experimentally the transition form factor in the time-like region of the momentum transfer~\cite{Landsberg:1985kjr,Landsberg:1985gaz}. Namely, the form factor can modify the lepton spectrum as compared with that obtained for point-like particles. 

Using the form factor in Eq. (\ref{eq:vmd_form}) and the differential rate in Eq. (\ref{eq:dgamma}), the total branching ratio for the Dalitz decay is obtained by integration:
\begin{equation}
\frac{\Gamma(K^* \to K \ell^+\ell^-) }{\Gamma(K^* \to K \gamma)}
=  \int_{4m_\ell^2}^{(M_K^*-M_K)^2} \frac{d\Gamma(K^* \to K \ell^+\ell^-)}{dq^2} dq^2.
\label{eq:totalBR}
\end{equation}

We perform the integration in Eq. (\ref{eq:totalBR}) separately for the dielectron ($\ell=e$) and dimuon ($\ell=\mu$) channels. The lower integration limit $4m_\mu^2$ for the muon channel significantly reduces the available phase space compared to the electron channel, leading to a strong suppression of the muon mode due to the $(1 - 4m_\ell^2/q^2)^{1/2}$ factor.

With the known radiative branching fraction and  the masses of the $K^*(892)$, leptons as well as kaons taken from the Particle Data Group (PDG)~\cite{ParticleDataGroup:2024cfk}, our calculation yields the following predictions for both charged and neutral $K^*$ decaying into kaon plus leptonic pair, which are summarized in Table~\ref{t2}, where the uncertainties are from the branching fractions of radiative decays.

\begin{center}
\begin{table}[h]
\caption{The estimated decay rates for $K^* \rightarrow K \ell^+\ell^-$, based on Eq.~(\ref{eq:dgamman}) using the single-pole approximation of Eq.~(\ref{eq:pole}) with $m_\rho$. The error on the decay rate is from the measured error of $K^*\rightarrow K \gamma$ which is used as normalization.}

\label{t2}
\begin{tabular}{c|c|c}\hline
  Decay mode  & $e^+e^-$ & $\mu^+\mu^-$ \\ \hline
$K^{*+} \rightarrow K^+ l^+ l^-$ & $(7.94\pm 0.73)\times 10^{-6}$ & $(2.35\pm 0.22)\times 10^{-7}$
\\  \hline
$K^{*0} \rightarrow K^0 l^+ l^-$ & $(1.99\pm 0.17)\times 10^{-5}$ & $(5.9\pm 0.5)\times 10^{-7}$
\\ \hline
\end{tabular}
\end{table}
\end{center}

The shape of $d\Gamma/dq^2$  shows a sharp rise near the low-$q^2$ threshold (especially for electrons).
This structure is a clear prediction of the single-pole VMD model and its experimental verification would be a significant test of the framework for strange mesons. Comparing our predicted spectra and branching fractions  with future experimental data will either validate this approximation for the $K^*K$ transition or point towards necessary refinements, such as including contributions from heavier vector resonances ($\omega$ or $\phi$) in a full multi-resonance VMD treatment, or applying alternative theoretical models ($e.g.$, dispersive approaches). 

\begin{center}
\begin{figure}[htbp]
\subfigure{
\includegraphics[width=0.4\textwidth, page=1]{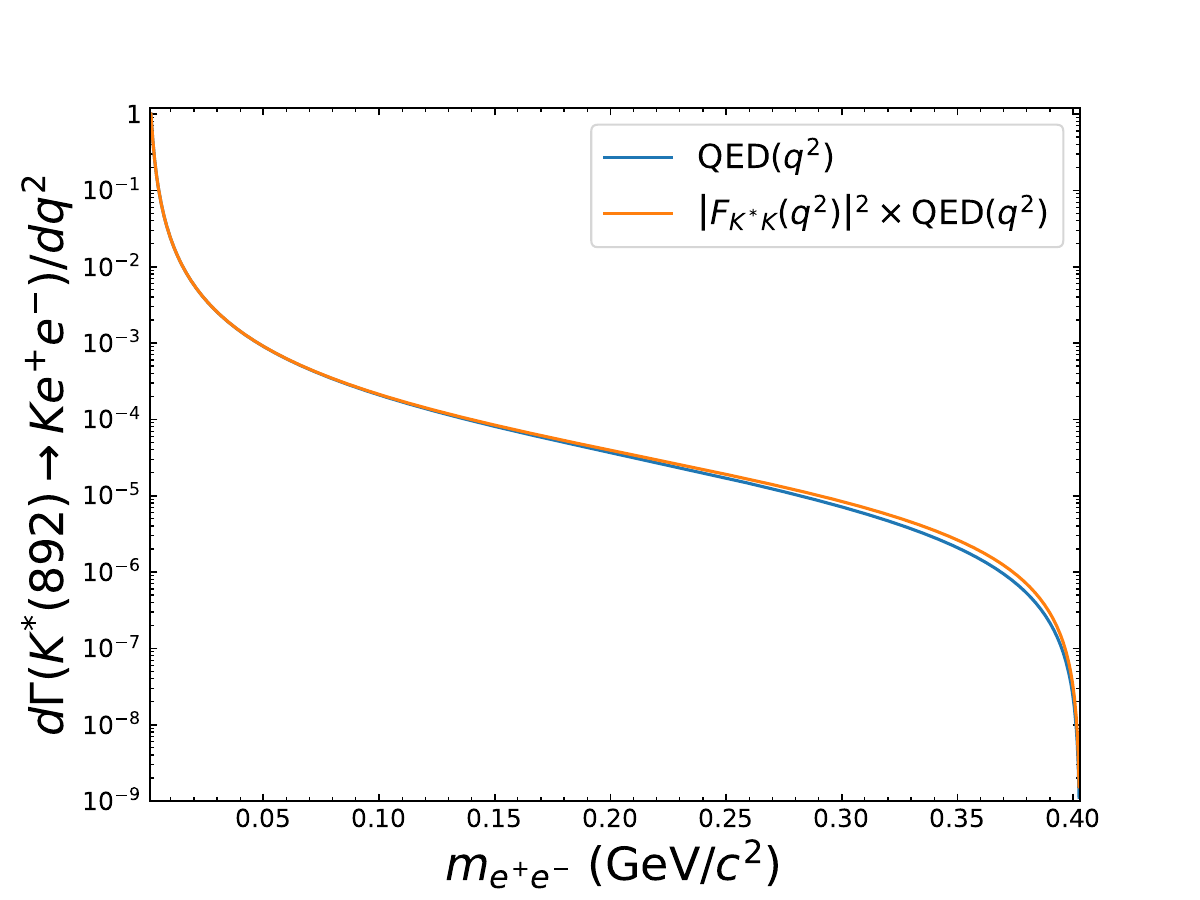}
}
\subfigure{
\includegraphics[width=0.4\textwidth, page=2]{V2Pll.pdf}
}
\caption{The differential
decay rate for $K^*(892) \rightarrow K  \ell^+ \ell^-$, where the solid curve
is for the pole mass taken as $\Lambda=m_{\rho}$.} \label{poledepen1}
\end{figure}
\end{center}

The decay channel $J/\psi \to K^*(892) K$, followed by the electromagnetic process $  K^*(892) \to  K \ell^+ \ell^- $, offers an intriguing opportunity to search for rare kaon resonance decays at flavor factories. The branching fraction for $J/\psi \to K^* K$ is approximately $\mathcal{O}(10^{-3}) $, while the branching fraction for $  K^*(892) \to e^+ e^- K $ is estimated to be $  10^{-5}$ . Therefore, the combined branching fraction for the full chain can be roughly evaluated to be $\mathcal{B}(J/\psi \to K^*(892)K \to K e^+ e^- K) \sim 10^{-8}$.

At the BESIII experiment, with a dataset of approximately 10 billion $J/\psi$ events~\cite{BESIII:2021cxx}, around 100 raw signal events are expected before accounting for detection efficiency and acceptance. Given an overall efficiency of about 20\%, this corresponds to roughly 20 reconstructed signal candidates. Benefiting from BESIII’s excellent lepton and kaon identification capabilities and good mass resolution, the potential for a first observation and branching fraction measurement becomes realistic. Nevertheless, substantial effort is required to control backgrounds from other $J/\psi$ decays that produce similar final states. Particularly, decays such as $J/\psi\to\gamma KK$, where the kaon pair originates from intermediate resonances like $f_0(1710)$ or $f_2(1270)$, can contribute significant backgrounds. Similar challenges were observed in Ref.~\cite{BESIII:2018ubj}, where the $K^*(892)$ signal appeared above a smooth but considerable background. Given the extremely small branching fractions of the muon channel ($\mathcal{O}(10^{-7})$), observing $K^*(892) \to K \mu^+\mu^-$ at BESIII is not feasible with current statistics. In contrast, the dielectron channel, with its $\mathcal{O}(10^{-5})$ branching fraction, offers realistic prospects for a first observation. Advanced analysis methods, including multivariate techniques and mass-sideband subtraction, will be essential to maximize sensitivity for the electron decay channel.

Looking forward to the Super Tau-Charm Facility (STCF)~\cite{Achasov:2023gey}, which is designed to collect samples exceeding $10^{12}$ $J/\psi$ events annually, the available statistics for such rare channels will improve by at least two orders of magnitude. 
This would allow for precision measurements of branching fractions and offer opportunities to test lepton-flavor universality and examine differential distributions in the decay.
With further improved detector capabilities at STCF---such as superior particle identification and acceptance---future sensitivity for $J/\psi \to K^*(892)(\to \ell^+\ell^- K) K  $ can be expected to increase dramatically, opening new windows for rare electromagnetic hadronic transitions and potential new physics searches.

\section{Dark photon sensitivity}

The Dalitz decay $K^*(892)\to \ell^+\ell^- K $ proceeds via an intermediate virtual photon, $\gamma^*$, converting into a lepton pair. If a kinetically mixed dark photon $A^\prime$ exists with mass $m_{A^\prime}$ below the kinematic endpoint
 ($m_{A^\prime} < m_{K*} -m_{K}\sim$ 398 MeV/$c^2$), it can be produced on shell in  $K^*(892) \to K A'$  and subsequently decay $A' \to \ell^+\ell^-$, appearing as a narrow peak over the smooth Dalitz continuum in the dilepton invariant-mass spectrum. The search strategy and the theoretical description closely parallel those laid out for $J/\psi$ Dalitz decays~\cite{Fu:2011yy,BESIII:2025xjh,He:2020jvj,BESIII:2018qzg,BESIII:2018aao,BESIII:2014dax}.

The key production width $\Gamma(K^* \to K A')$ is related to the real photon width by the mixing,
\begin{widetext}
\begin{equation}
\mathcal{B}(K^* \to K A') = \varepsilon^2 \cdot \frac{ \lambda^{3/2}(m_{K^*}^2, m_K^2, m_{A'}^2) }{\lambda^{3/2}(m_{K^*}^2, m_K^2, 0)} \cdot |F_{K^*K}(m_{A'}^2)|^2 \cdot \mathcal{B}(K^*\rightarrow \gamma K)\cdot \mathcal{B}(A'\rightarrow \ell^+\ell^-),
\label{eq:prod_width}
\end{equation}
\end{widetext}
where $\lambda(x,y,z) = x^2 + y^2 + z^2 - 2xy - 2yz - 2zx$, $m_l$ is the lepton mass, and $m_{K^*}$, $m_K$ are the masses of the $K^*$ and $K$ mesons, respectively. The production branching fraction $\mathcal{B}(K^* \to K A')$ is proportional to $\varepsilon^2$, modulated by the form factor squared evaluated at the dark photon mass. The visible branching fraction for the full chain $K^* \to K A'(\to \ell^+\ell^-)$ further includes the factor $\mathcal{B}(A' \to \ell^+\ell^-)$, which depends on the dark photon decay model; in our sensitivity estimate we conservatively set $\mathcal{B}(A'\to\ell^+\ell^-)=1$ for maximal visibility.

The total observed dilepton invariant mass spectrum would be the sum of the smooth SM Dalitz continuum (background, $B$) and the narrow dark photon resonance (signal, $S$):
\begin{widetext}
\begin{equation}
\left.\frac{dN}{dq^2}\right|_{\text{obs}} = N_{K^*} \left[ \frac{d\mathcal{B}(K^*\to\ell^+\ell^-K)}{dq^2} + \frac{d\mathcal{B}(K^*\to A^\prime K)}{dq^2} \right] \cdot \epsilon_{\text{det}}(q^2),
\label{nobs}
\end{equation}
\end{widetext}
where $N_{K^*}$ is the number of produced $K^*$ mesons and $\epsilon_{\text{det}}$ is the detection efficiency.

For $m_{A'}$ significantly below $(M_{K^*} - M_K)$, the signal appears as a narrow peak. Its visibility is determined by the signal-to-background ratio $S/B$ in the resonant mass window. The resonant nature makes the search highly sensitive even to very small $\varepsilon$. In this case, we could estimate  the experimental sensitivity to the kinetic mixing parameter $\varepsilon$ by considering a simplified analysis.  

\begin{figure}
    \centering
    \includegraphics[width=0.4\textwidth]{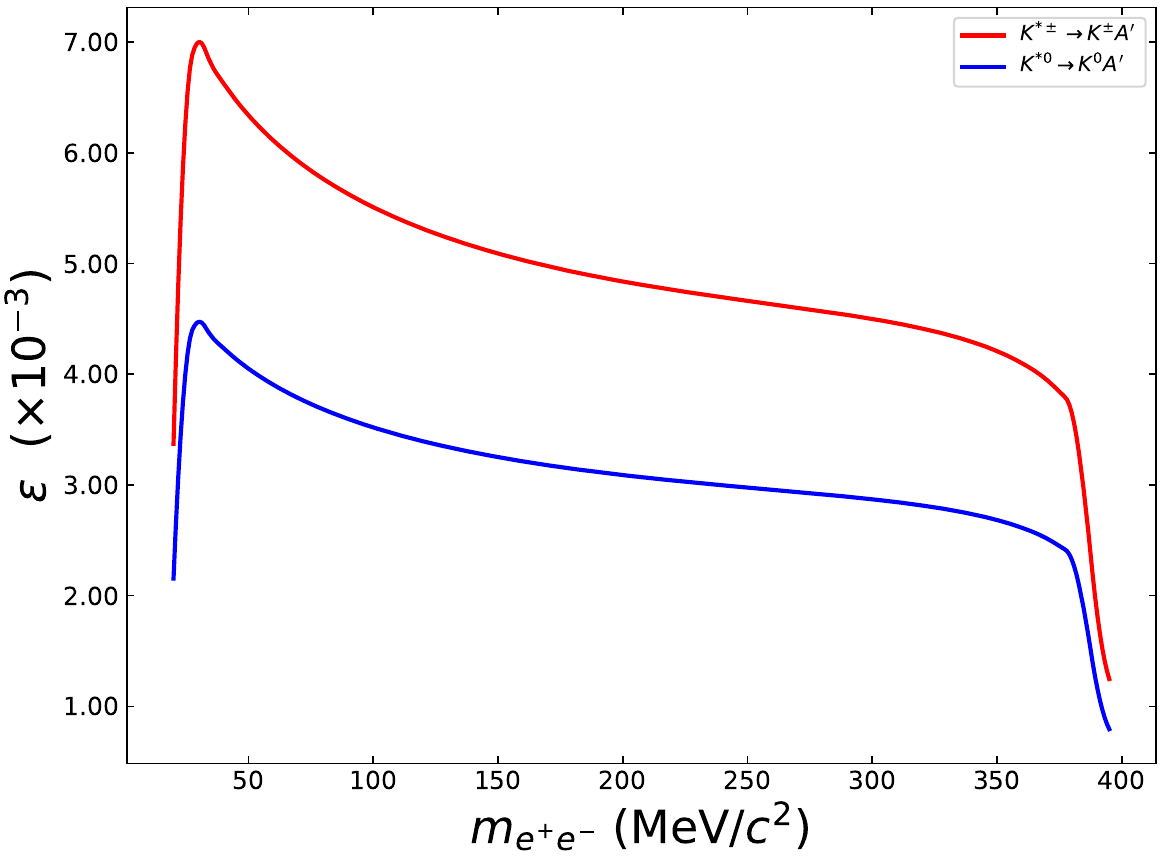}
    \caption{The sensitivity of the coupling constant $\varepsilon$ of $A^\prime$ estimated with the 10 billion $J/\psi$ events at BESIII experiment . }
    \label{fig:up}
\end{figure}

As shown in Eq.~(\ref{nobs}), for a given number of reconstructed  decays, we consider the contributions from both Dalitz decay $K^*\to\ell^+\ell^-K$ and the possible dark photon contribution from $K^*\to A^\prime K$.
Assuming that the background events originate only from the Dalitz decay of $K^*\to\ell^+\ell^-K$,  the toy MC samples were generated in accordance with the branching fractions~\cite{ParticleDataGroup:2024cfk} of $J/\psi\to K^{*\pm}K^{\mp}(K^{*\pm}\to\ell^+\ell^- K^\pm)$ and $J/\psi\to K^{*0}\bar{K^{0}}(K^{*0}\to\ell^+\ell^- K^0)$  and then normalized to the 10 billion $J/\psi$ events at BESIII experiment. By assuming a narrow width for the dark photon, its signal shape is modeled with a Gaussian function. The upper limit at a 90\% confidence level is then estimated using Pearson's $\chi^2$ test. Take $K^{*+}(892)\rightarrow KA^\prime$ for example,  the sensitivity to the dark photon coupling constant $\varepsilon$ is estimated to reach the order of $10^{-3}$ at 90\% confidence level, as shown  in Fig.~\ref{fig:up},  based on the 10 billion $J/\psi$ events accumulated at BESIII experiment.

Due to the low production of $K^*(892)$, the sensitivity for $K^* \to K \ell^+\ell^-$ may be not compatible with existing limits from other processes like $\pi^0 \to \gamma A'$, $\eta \to \gamma A'$, and direct $e^+e^- \to \gamma A'$ searches. However, the $K^*$ channel offers unique coverage in specific mass regions, particularly for $m_{A'}$ between the $\phi$ mass and the kinematic threshold $\sim 400$ MeV/c$^2$, where other meson decays may have less sensitivity. The muon channel, despite its lower rate, provides crucial constraints in regions where the dielectron channel might face large backgrounds from photon conversions. A combined analysis of both channels would maximize discovery potential. The observation of a statistically significant narrow peak would constitute a smoking-gun signature for a dark photon. The measured mass $m_{A'}$ and the peak yield would provide key parameters for dark sector model building. 

\section{Summary }
In this work, we  presented a study of the rare Dalitz decay $K^*(892) \to K\ell^+\ell^-$, addressing its dual role as a probe of hadronic structure and a promising channel for dark photon searches.
We have calculated the branching fractions for both electron and muon final states using the Vector Meson Dominance model, providing the first dedicated predictions at the level of $\mathcal{O}(10^{-5})$ and $\mathcal{O}(10^{-7})$, respectively. These results establish essential expectations for future experimental measurements and will test the applicability of VMD to strange vector meson transitions.

Furthermore, we have analyzed the sensitivity of this decay to a hypothetical dark photon. We demonstrated how a narrow resonance would appear in the dilepton mass spectrum and derived projected exclusion limits on the kinetic mixing parameter $\varepsilon$ across the accessible mass range. Our analysis shows that high-statistics experiments at facilities like BESIII, LHCb, or a future STCF can achieve competitive sensitivity, particularly in mass regions complementary to other search channels.

This study underscores the significant untapped potential of $K^*(892) \to K\ell^+\ell^-$. Precision measurements of this decay will advance our understanding of QCD dynamics in the strange sector, while simultaneously conducting a sensitive, direct search for dark sector particles. We strongly encourage experimental collaborations to undertake dedicated analyses of this channel, as it promises rich dividends in both hadron physics and the quest for new physics beyond the Standard Model.

\section{Acknowledgement}
This work is supported  by the National Natural Science Foundation of China under contract No. 12225509; This work is in part supported in by  Henan Provincial Natural Science Foundation for Young Scientists  under contract No. 252300421773.


\bibliographystyle{apsrev4-2}
\bibliography{draft-ref}
\end{document}